\documentclass[aps,prl,twocolumn,showpacs,superscriptaddress,floatfix]{revtex4}
\usepackage{graphicx}
\usepackage{bm}

\begin{document}
\newcommand{\beq}{\begin{equation}}
\newcommand{\eeq}{\end{equation}}

\title{Merging of single-particle levels and non-Fermi-liquid behavior
of finite Fermi systems}

\author{V.~A.~Khodel}
\affiliation{ Russian Research Centre Kurchatov
Institute, Moscow, 123182, Russia}
\affiliation{ McDonnell Center for the Space Sciences and
Department of Physics, Washington University,
St.~Louis, MO 63130, USA}

\author{J.~W.~Clark}
\affiliation{ McDonnell Center for the Space Sciences and
Department of Physics, Washington University,
St.~Louis, MO 63130, USA}

\author{Haochen Li}
\affiliation{ McDonnell Center for the Space Sciences and
Department of Physics, Washington University,
St.~Louis, MO 63130, USA}

\author{M.~V.~Zverev}
\affiliation{ Russian Research Centre Kurchatov
Institute, Moscow, 123182, Russia}

\date{\today}

\begin{abstract}
We examine the problem of finite Fermi systems having a degenerate
single-particle spectrum and show that the Landau approach, applied to
such a system, admits the possibility of merging single-particle
levels.  It is demonstrated that the opportunity for this behavior
is widespread in quantum many-body systems.  The salient feature
of the phenomenon is the occurrence of nonintegral quasiparticle
occupation numbers, leading to a radical alteration of the standard
quasiparticle picture.  Implications of this alteration are
considered for nuclear, atomic, and solid-state systems.
\end{abstract}

\maketitle

Landau Fermi-liquid theory \cite{lan} (FL) is recognized as one of
the foundation stones of our understanding of the properties
of condensed matter.  When adapted to finite Fermi systems
in nuclear and atomic physics, the standard FL quasiparticle
picture prescribes that the total angular momenta of the ground
states of odd nuclei must be carried by a single quasiparticle,
and that the electronic configurations of ions of elements
of the periodic table must repeat those of preceding atoms.
Its iconic stature notwithstanding, the standard FL picture
sometimes fails.  Currently, the origin of non-Fermi-liquid
(NFL) behavior of homogeneous Fermi systems is one of the central
issues of condensed-matter physics.  Inhomogeneous, finite
Fermi systems, although unburdened by the damping of
single-particle (sp) excitations cited as the source of failure
of the Landau quasiparticle picture \cite{pines}, are also known to
exhibit violations of FL theory.  For example, the total angular
momenta of the ground states of many odd-$A$ nuclear isotopes
in the transition region cannot be attributed to a sp state,
and the electronic configurations of elements not belonging to
principal groups differ from those expected in FL theory.

Here we focus attention on finite Fermi systems whose sp
spectrum $\epsilon_\lambda$ has a degeneracy and show that,
rather surprisingly, NFL behavior can arise within the
Landau approach \cite{lan} itself, with the merging of
sp levels lying on opposite sides of the Fermi surface.
This phenomenon, almost never addressed in condensed-matter
physics, stems from the variation of sp energies under
change of the occupation of the last unfilled sp level due
to the interaction between quasiparticles.  In the systems
to be studied, the energetic distance between a sp level being
filled and the nearest empty level shrinks progressively as
the former level is filled, leading to a crossing
of the levels in cases where standard FL theory is
obeyed.  However, in the case of interest, the levels do
not cross one another; instead, they {\it merge}.  As will be
seen, a primary condition for merging to occur is that the
Landau-Migdal interaction function $f$ is repulsive in
coordinate space, which holds for the interactions between
particles of the same kind in the nuclear interior and
for electron-electron interactions in atoms.

We begin the analysis of this phenomenon by considering
two neutron levels in an open shell of of a schematic
model of a spherical nucleus of mass number $A$ and radius
$R=r_0A^{1/3}$.  The levels are denoted by $0$ and $+$, in order of
increasing energy.  As usual, the sp energies $\varepsilon_\lambda$
and wave functions $\varphi_{\lambda}({\bf r})=R_{nl}(r)\Phi_{jlm}({\bf n})$
are solutions of equation $[p^2/2M+\Sigma({\bf r},{\bf p})]\varphi_{\lambda}({\bf r}) =
\epsilon_{\lambda}\varphi_{\lambda}({\bf r})$, where $\Sigma$ stands
for the quasiparticle self-energy. In a spherical nucleus with
even numbers of neutrons and protons, which has zero total angular
momentum in its ground state due to pairing correlations, the sp
energies $\epsilon_{\lambda}$ are independent of the magnetic
quantum number $m$.  We suppose that the orbital angular momenta of
levels $0$ and $+$ obey $l_0\neq l_+\gg 1$ and follow the variation of
the distance between these levels, as $N\gg 1$ neutrons are
added to the level $0$, changing the density $\rho(r)$ by
$\delta\rho(r)=NR^2_{n_0l_0}(r)/4\pi$.  We neglect self-interaction
corrections and retain only a major, spin- and momentum-independent
part $V(r)$ of the self-energy $\Sigma$ and a primary, $\delta(r)$-like
portion of the Landau-Migdal interaction function $f$. In this case,
the FL relation between the self-energy and the density $\rho$ that is
responsible for the variation of $\epsilon_{\lambda}$ simplifies
to \cite{migdal,schuck}
\begin{equation}
  \delta V(r)=f[\rho(r)]\, \delta\rho(r) \  .
\label{mig}
\end{equation}
The interaction matrix elements
\begin{equation}
f_{kk'}=\int R_k^2(r)f\left[\rho(r)\right]R_{k'}^2(r){r^2dr\over 4\pi}\
\label{mel}
\end{equation}
are assigned values $f_{00}=u$, $f_{++}=v$, and $f_{0+}=w$.
In a semiclassical approximation where $R_k(r)\sim r^{-1}R^{-1/2}\cos\int p_k(r)dr$, one has $u\simeq v\simeq 3w/2$.

Based on these assumptions, the dimensionless energy shifts
$\xi_k(N)=\left[\epsilon_k(N)-\epsilon_k(0)\right]/D$
are given by
\begin{equation}
\xi_0(N)=n_0U\ \quad {\rm and} \quad \xi_+(N)=n_0W \  ,
\label{en1}
\end{equation}
where $n_k=N_k/(2j_k+1)$ is the occupation number of level $k$,
$D$ is the initial distance between levels $+$ and 0,
$U=u(2j_0+1)/D$, and $W=w(2j_0+1)/D$.  We have neglected
second-order corrections that have little effect on the
results since they are almost independent of the sp quantum
numbers.

According to Eqs.~(\ref{en1}), the dimensionless distance
$d(N)=\left[\epsilon_+(N)-\epsilon_0(N)\right]/D=1+\xi_+(N)-\xi_0(N)$
changes sign at $N_c=D/(u-w)$. This always occurs before filling
of the level $0$ is complete if the distance $D$ is rather small.
For $N$ above some critical value $N_c$, Eqs.~(\ref{en1}) become
inapplicable, since the pattern of orbital filling must change.
In the standard FL picture, which allows only for crossing of sp levels,
all the added $N>N_c$ quasiparticles must settle into the empty sp
level $+$, which entails a change of the relevant sp energies. The sign
of the difference $\epsilon_r(N)=\delta\epsilon_+(N)-\delta\epsilon_0(N)$
of the shifts of the sp energies $\epsilon_+$ and $\epsilon_0$ due
to total migration of the quasiparticles from level $0$ into
level $+$ is crucial.  With the help of Eq.~(\ref{mig}), it is
seen that
\begin{equation}
\epsilon_r(N)=N(u-2w+v)\  .
\label{crit}
\end{equation}
In our model, $\epsilon_r(N)\simeq 2Nu/3$ is positive.  The positivity
of this key quantity means that level $+$ remains above
level $0$ upon the migration.  The standard FL scenario must then
encounter a catastrophe at $N>N_c$: on the one hand, quasiparticles
must leave level $0$; on the other, their total migration into
level $+$ is prohibited. To end the deadlock, {\it both} levels must be
partially occupied, in contradistinction to FL theory where {\it
one and only one} level can be partially occupied.  Such dual
partial occupation is possible only if the sp energies $\epsilon_0$
and $\epsilon_+$ coincide with the chemical potential $\mu$.
If so, at $N>N_c$, Eqs.~(\ref{en1}) becomes
\begin{equation}
\epsilon_0(0)+N_0u+N_+w = \epsilon_+(0)+N_0w+N_+v\,,
\label{en2}
\end{equation}
which has the solution
\begin{equation}
N_+= (N-N_c)(u-w)(u+v-2w)^{-1} \ .
\label{crit1}
\end{equation}

Significantly, a resolution of the dilemma has been found within the Landau
approach itself.  To understand the consistency of this resolution,
recall that the occupation numbers of Landau quasiparticles
are given by \cite{lan}
$n_{\lambda}(T)=[1+e^{(\epsilon_{\lambda}-\mu)/T}]^{-1}$.
At $T=0$, this formula guarantees that occupation numbers are
restricted to the values 0 and 1, but only for those sp
levels with energies $\epsilon_{\lambda}\neq\mu$; otherwise
the index of the exponent is uncertain.  As seen above, the
merging of two (or more) sp levels drives the levels exactly
to the Fermi surface. Solution of equations of merging such
as (\ref{en2}) removes the residual uncertainty at the expense of
introducing fractional occupation numbers $n_{\lambda}$,
violating what would appear to be an elementary truth of
FL theory, but in fact maintaining consistency.  The sp Green
function in the presence of merging has the familiar form
\begin{equation}
G({\bf r},{\bf r}',\epsilon)=
  \sum_{k=0,+}
\left({1-n_k\over \epsilon{-}\epsilon_k{+}i\delta}+
{n_k\over \epsilon{-}\epsilon_k{-}i\delta}\right)
  \varphi_k({\bf r})\,\varphi_k^*({\bf r}') \ ,
\label{green}
\end{equation}
but the occupation numbers $n_k$ of the merging sp levels
become fractional.
\begin{figure}[ht]
\includegraphics[width=0.35\textwidth,height=0.35\textwidth]{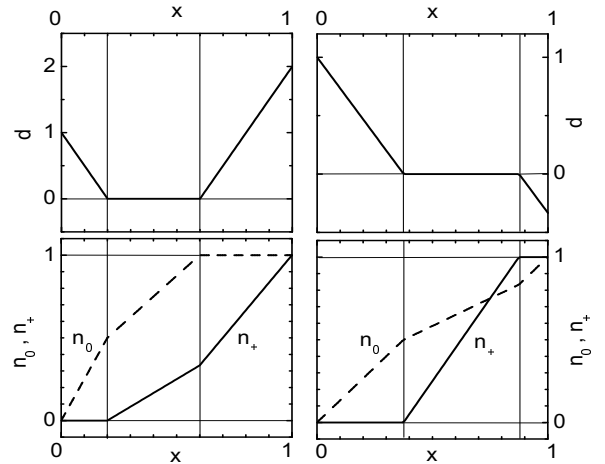}
\caption{Top panels:
Dimensionless distance $d=(\epsilon_+-\epsilon_0)/D$
between levels $+$ and $0$ as a function of the ratio
$x=N/(2j_0+2j_++2)$. Lower panels:
Occupation numbers $n_k$ for levels $0$ and $+$. Input parameters:
$U=V=3, W=1$. For the left column, the ratio
$r\equiv (2j_0+1)/(2j_++1)=2/3$; for the right, $r=3$.}
\label{fig:fig1}
\end{figure}

Results from numerical calculations are plotted in Fig.~\ref{fig:fig1},
which consists of two columns, each made up of two plots.  The
upper panels show the dimensionless ratio
$d(x)=\left[\epsilon_+(x)-\epsilon_0(x)\right]/D$ versus $x=N/(2j_0+2j_++2)$,
which runs from 0 to 1.  The lower panels give the occupation numbers
$n_+(x)$ and $n_0(x)$, which, in the range of $x$ where
$d(x)=0$, behave as $n_+(x)=x(1+r)/2-r/(2(U-W))$ and
$n_0(x)=x(1+r)/(2r)+1/(2(U-W))$, with $r=(2j_0+1)/(2j_++1)$.

In the variable $x$, the model exhibits three distinct regimes.
In two of them, $d\neq 0$ and the standard FL picture holds.
It fails in the third region, where $d=0$ and integration of
Eq.~(\ref{green}) over $\epsilon$ yields the density
$\rho({\bf r})=\sum_{k=0,+} n_k\varphi_k({\bf r})\varphi_k^*({\bf r})$
for the added quasiparticles.  This result cannot be attributed
to a single sp level, implying that well-defined sp excitations
in the familiar Landau sense {\it no longer exist}.  Passage
through the three regimes can be regarded as a second-order
phase transition, with the occupation number $n_+(x)$ treated as
an order parameter.

The two sp levels remain merged until one of them is completely
filled. If the level $0$ fills first, as in the left column of
Fig.~\ref{fig:fig1}, the episode of merging is ended by repulsion of
the two levels, as if they possess the {\it same symmetry} --
despite the fact that in the open shell, they always have
{\it different} symmetries. In another case where level
$+$ becomes fully occupied before level 0, as in the right
column, the distance $d(x)$ becomes negative, and the two
levels just cross each other at this point.

In the nuclear many-body problem, both types of sp level degeneracy --
either initially present or arising in the scenario described above --
are lifted when pairing correlations are explicitly involved
\cite{belyaev,migdal,schuck}.  In atomic nuclei, realistic pairing
forces are so weak that in nuclei with closed shells, pairing
correlations are completely suppressed, with the effect that
the nuclear pairing energy becomes a part of a shell correction
to the Bethe-Weisz\"acker liquid-drop formula.  A model calculation
\cite{condmat} shows that merging of even two sp levels approximately
doubles the gap value and drastically increases the pairing energy.
More realistic calculations are needed to determine whether this
enhancement is enough to promote the existence of a new minimum
in the ground-state energy functional of superheavy nuclei.

We turn now to the merging of sp levels in atoms.  In fact, it
was in atomic physics that one of the first models of NFL behavior
involving nonintegral occupation numbers was developed by
Slater et al.~\cite{slater}.  This model is based the observation
that results of Hartree-Fock (HF) calculations for atoms of
intermediate groups are often improved if, in an extended HF
energy functional, contributions of two leading Slater configurations
are included additively with factors $0<x<1$ and $1-x$.
The best choice for $x$ is found from a minimization procedure,
yielding equations similar to Eqs.~(\ref{en2}), which naturally
entail fractional occupation numbers $n_{\lambda}$.  As we have
already seen within the framework of the non-perturbative Landau
approach, this feature is not a prerogative or artifact of the
HF method: nonintegral occupation numbers may legitimately come
into play provided the sign of the key quantity $\delta\epsilon_r$ from
Eq.~(\ref{crit}) is positive, and, hence, the criterion for
merging of sp levels is met.  It therefore comes as no surprise
that the conditions for merging of sp levels are satisfied in
certain strongly correlated Fermi systems \cite{ks,physrep} for
which the HF method is inapplicable.

Two circumstances complicate the analysis of merging electron sp
levels in atoms.  First, the sp energies $\epsilon_{njlm}$ cease
to be $m$-independent due to the absence of Cooper pairing.
Difficulties stemming from this fact can be avoided if, following
Ref.~\onlinecite{slater}, one tracks the center-of-gravity
$\epsilon_k^{\rm o}= \sum_m \epsilon_{km}/(2j_k+1)$ energies of
levels, rather than individual $m$-levels or a band.  Then one
only has to deal with the spherically symmetric part of
$\delta\rho({\bf r})$, as in the nuclear problem.  Second, the
self-energy $\Sigma$ has a nonlocal character due to the presence
of long-range Coulomb interactions.  In another respect, the treatment
of merging sp levels in atomic systems is much simpler than in the
nuclear case, because the spacing parameter $r_s=r_0/a_B$ is less than
unity ($r_0$ being the radius of the volume per electron and
$a_B$, the Bohr radius).  This implies that correlation
contributions to the electron-electron interaction function
$f_{ee}$ are rather small compared to exchange \cite{ashag}, so
that $f_{ee}$ takes the Hartree-Fock form.  Carrying out the
same operations that led to the necessary condition (\ref{crit}), we
readily arrive at the corresponding condition
\beq
f_{nl}^{nl}+f_{n'l'}^{n'l'}-2f_{nl}^{n'l'}>0
\label{critc}
\eeq
for the merging of two electron sp levels with quantum numbers
$n,l$ and $n',l'$.  Introducing ${\cal R}_{nl}(r)=rR_{nl}(r)$,
the interaction matrix elements are constructed as
$$
f_{nl}^{n'l'}=e^2\int[ {\cal R}_{nl}^2(r_1){\cal R}^2_{n'l'}(r_2)
$$
\beq
  -
  (2j+1)^{-1}{\cal R}_{nl}(r_1){\cal R}_{n'l'}(r_1)
             {\cal R}_{nl}(r_2){\cal R}_{n'l'}(r_2)]{1\over r_>}
                                                       dr_1dr_2\,,
\label{melc}
\eeq
where $r_>$ is the greater of $r_1$, $r_2$.  In obtaining Eq.~(\ref{melc}),
we have neglected insignificant contributions to the exchange part of
$f_{ee}$ coming from multipole moments created by electrons moving in
the open shell.  Semiclassical estimates along the same lines as before
confirm that the nondiagonal matrix elements of $f_{ee}$ are of much
smaller size than the diagonal ones.  We may then conclude that the
difference on the l.h.s.\ of inequality (\ref{critc}) is positive
independently of the quantum numbers, so that the necessary condition
for merging of the sp levels is met.

It is instructive to compare the values of the basic parameter $N_c$
that governs the merging phenomenon in nuclear and atomic systems.
In the nuclear problem, the neutron-neutron interaction
is characterized by the dimensionless constant \cite{migdal}
$F_{NN}=f_{NN}p_FM/\pi^2\simeq 1$; from Eq.~(\ref{mel}) one
then obtains $u \simeq \epsilon^0_F/A$, where $\epsilon^0_F=p^2_F/2M$
is the Fermi energy.  On the other hand, the distance $D$ between
sp levels in spherical nuclei with closed shells is of order of
$\epsilon^0_F/A^{2/3}$.  Consequently, the critical particle number
$N_c=D/(u-w)$ must be of order of $ A^{1/3}$, which in turn means that
not every pair of nuclear sp levels adjacent to the Fermi surface
is susceptible to merging.

The situation in heavy atoms is quite different: the diagonal matrix
elements of the electron-electron interaction in the open shell
are of order several eV, markedly enhanced in the event
$f$-orbital collapse, as occurs in rare-earth and transuranic
elements \cite{gm,griffin,band,kikoin,lanl}.  The distance $D
$ between the sp levels adjacent to the Fermi surface in atoms
with closed electron shells is known to be some 1--2 eV as well,
so the critical number $N_c$ in atoms is {\it of order of unity}.
Consequently, in elements with nuclear charge $Z>20$, the
sp Green function has the form (\ref{green}) -- except for elements
of principal groups, where the standard FL picture still holds.
Furthermore, when $N$ substantially exceeds $N_c$, as is the
case of many rare-earth and transuranic elements, merging of
levels triggered by collapse of the $f$ orbitals inescapably
involves most of the sp levels in the open shell.  Fractional
occupation numbers become a necessity, with the result that these
elements lose their chemical individuality, a well-known feature
of the sequence of rare-earth elements.  The veracity of these
inferences can be tested by means of precise measurements of the
difference $\sigma(Z+1)-\sigma(Z)$ between cross sections for
elastic scattering of charged particles by rare-earth or
transuranic atoms with atomic numbers differing by unity.  In
the theory of Goeppert-Mayer \cite{gm}, this difference is
directly expressed in terms of the single 4$f$ wave function,
whereas if merging occurs, the density change involves all the
merged sp levels.

According to the above argument, in systems devoid of pairing
the centers of merged sp levels ``get stuck'' at the Fermi surface.
We observe that this could provide a simple mechanism for pinning
narrow bands in solids to the Fermi surface.  To exemplify
this point, consider a model in which the electron sp spectrum,
calculated in local-density approximation (LDA), is exhausted
by (i) a wide band that disperses through the Fermi surface,
and (ii) a narrow band, placed below the Fermi surface at a
distance $D_n$.  Turning on the electron-electron interactions
produces a change of the sp energies in accordance with the
Landau equation
\beq
 \epsilon({\bf p})=\epsilon_{\mbox{\scriptsize LDA}}({{\bf p}})
+\int f({\bf p},{\bf p}_1)  n(\epsilon({\bf p}_1))d^3p_1/(2\pi)^3  \   .
\label{lansp}
\eeq
To proceed, we assume that only matrix elements
$f^{(n)}({\bf p},{\bf p}_1)$ of the interaction function
$f$ referring to the narrow band are significant, the others
being negligible.  If the shift $\delta\epsilon^{(n)}$ in the
location of the narrow band due to switching on the
intraband interactions exceeds the distance $D_n$, then the
standard FL scenario calls for the narrow band to be completely
emptied; but then $\delta\epsilon_n$ must vanish, and
quasiparticles are obliged to return.  To eliminate this
mismatch, only a fraction of the particles leave the
narrow band, in just the right proportion to equalize
the band chemical potentials.  The feedback mechanism
we have described positions the narrow band exactly at
the Fermi surface, resolving a long-standing problem in
the LDA scheme.

It is worth noting that the bare narrow-band group velocity,
proportional to the corresponding bandwidth $W^{(n)}$, is rather
small, whereas the matrix elements $f^{(n)}({\bf p},{\bf p}_1)$,
which do not contain this factor, are not suppressed.
Consequently, the group velocity might change its sign when
the interaction correction is taken into account, giving rise
to the previously studied phenomenon known as fermion
condensation, which involves wholesale mergence of
sp levels in homogeneous Fermi fluids \cite{ks,physrep,noz}.
In spite of evident commonalities, there is a crucial
difference between the conditions for the ``level-mergence''
phenomenon in homogeneous Fermi liquids and in finite Fermi
systems with degenerate sp levels.  In the former, the
presence of a significant velocity-dependent component in
the interaction function $f$ is needed to promote fermion
condensation, while in the latter, sp levels can merge
even if $f$ is momentum-independent. The reason for this difference
is simple. In the homogeneous case, the matrix elements $u$, $v$,
and $w$ are equal to each other, implying zero energy gain due
to the rearrangement when velocity-dependent forces are absent.
The same is seen from Eq.~(\ref{lansp});
the group velocity, whose sign determines whether the Landau FL
state is stable, is unaffected by the momentum-independent part
of $f$.

The merging of sp levels quite often violates a symmetry inherent
in the initial ground state.  In nuclei, for example,
the members of pairs of sp levels with quantum numbers $n,l$
and $n\pm1, l\pm2$ lie quite close to each other.  As the lower
of the two levels is being filled, the distance from its neighbor
shrinks, resulting in an increase of the nuclear quadrupole moment.
If merging occurs, spherical symmetry is broken.  This mechanism
is presumably responsible for the occurrence of islands of nuclear
deformation beyond the mainland of rare-earth elements.  Merging
can also promote the enhancement of parity violation effects
in nuclei and atoms.  These issues will be addressed in a future
article.

In conclusion, our exploration of the mergence of single-particle
levels in finite Fermi systems exhibiting degeneracy reveals
that if the merging of two or more sp levels occurs, the
ground state experiences a rearrangement that introduces a
a multitude of quasiparticle terms, endowing it with
a more complex character, as in the comparison of a chorus
with a dominant soloist.

We thank K.~Kikoin, E.~Saperstein, and V.~Yakovenko
for fruitful discussions. This research was supported by
Grant No.~NS-8756.2006.2 from the Russian Ministry of
Education and Science and by the McDonnell Center for the
Space Sciences.

\end{document}